\documentclass[reprint,amsmath,amssymb,aps,superscriptaddress]{revtex4-2}
\usepackage{CJK}
\usepackage{graphicx} % Required for inserting images
\usepackage{dcolumn}
\usepackage{bm}
\usepackage[dvipsnames]{xcolor}
\begin{document}
%\begin{CJK*}{GB}{}
\title{Optomechanical Backaction in the Bistable Regime}
\author{L. F. Deeg}% \orcidlink{0000-0001-9783-1975}}
\email{lukas.deeg@uibk.ac.at}
\affiliation{Institute for Quantum Optics and Quantum Information, Austrian Academy of Sciences, 6020 Innsbruck, Austria}
\affiliation{University of Innsbruck, Institute for Experimental Physics, 6020 Innsbruck, Austria}
\author{D. Zoepfl}
\affiliation{Institute for Quantum Optics and Quantum Information, Austrian Academy of Sciences, 6020 Innsbruck, Austria}
\affiliation{University of Innsbruck, Institute for Experimental Physics, 6020 Innsbruck, Austria}
\author{N. Diaz-Naufal}
\affiliation{Dahlem Center for Complex Quantum Systems and Fachbereich Physik, Freie Universit\"{a}t Berlin, 14195 Berlin, Germany}
\author{M. L. Juan}% \orcidlink{0000-0002-2740-8001}}
\affiliation{Institut Quantique and D\'epartement de Physique, Universit\'e de Sherbrooke, Sherbrooke, Qu\'ebec, J1K 2R1, Canada}
\author{A. Metelmann}% \orcidlink{0000-0001-9783-1975}}
\affiliation{Dahlem Center for Complex Quantum Systems and Fachbereich Physik, Freie Universit\"{a}t Berlin, 14195 Berlin, Germany}
\affiliation{Institute for Theory of Condensed Matter and Institute for Quantum Materials and Technology, Karlsruhe Institute of Technology, 76131 Karlsruhe, Germany} 
\affiliation{Institut de Science et d’Ingénierie Supramoléculaires (ISIS, UMR7006), University of Strasbourg and CNRS}

\author{G. Kirchmair}% \orcidlink{0000-0003-4286-8839}}
\email{gerhard.kirchmair@uibk.ac.at}
\affiliation{Institute for Quantum Optics and Quantum Information, Austrian Academy of Sciences, 6020 Innsbruck, Austria}
\affiliation{University of Innsbruck, Institute for Experimental Physics, 6020 Innsbruck, Austria}

\date{\today}

\begin{abstract}
     With a variety of realizations, optomechanics utilizes its light-matter interaction to test fundamental physics. By coupling the phonons of a mechanical resonator to the photons in a high-quality cavity, control of increasingly macroscopic objects has become feasible. In such systems, state manipulation of the mechanical mode is achieved by driving the cavity. To be able to achieve high drive powers the system is typically designed such that it remains in a linear response regime when driven. A nonlinear response, and especially bistability, in a driven cavity is often considered detrimental to cooling and state preparation in optomechanical systems and is avoided in experiments. Here we show that with an intrinsic nonlinear cavity backaction cooling of a mechanical resonator is feasible operating deep within the nonlinear regime of the cavity. With our theory taking the nonlinearity into account, precise predictions on backaction cooling can be achieved even with a cavity beyond the bifurcation point, where the cavity photon number spectrum starts to deviate from a typical Lorentzian shape.

\end{abstract}
\maketitle
Preparing a mechanical mode near its motional ground state has become a crucial task for fundamental physics experiments and quantum sensing applications \cite{aspelmeyerCavityOptomechanics2014}. However, even at cryogenic temperatures massive mechanical resonators reside in highly populated phonon states. This makes additional cooling necessary to bring them toward their motional ground state. Utilizing the interaction between the photon mode of a cavity and the phonon mode of a mechanical resonator, cooling of the mechanics can be achieved by feedback \cite{manciniOptomechanicalCoolingMacroscopic1998b,rossiMeasurementbasedQuantumControl2018b,tebbenjohannsMotionalSidebandAsymmetry2020} and sideband cooling \cite{marquardtQuantumTheoryCavityAssisted2007a,wilson-raeTheoryGroundState2007a}. For the latter, cooling can be achieved when driving the cavity below the resonance frequency, where we have to distinguish two cases. In the first case, the setup resides within the bad cavity regime when the cavity linewidth exceeds the mechanical frequency ($\kappa > \omega_m$). There the achievable cooling is limited by unwanted backaction heating \cite{aspelmeyerCavityOptomechanics2014}. In the opposite case, the so-called resolved sideband regime ($ \kappa< \omega_m$), ground state cooling can be reached as shown for various systems (e.g., systems based on microwave cavities) \cite{teufelSidebandCoolingMicromechanical2011b,chanLaserCoolingNanomechanical2011b}. 

However, at high photon numbers such systems feature a nonlinear behavior causing the linearization of the standard optomechanical theory to fail to describe the optomechanical interaction \cite{marquardtQuantumTheoryCavityAssisted2007a,safavi-naeiniLaserNoiseCavityoptomechanical2013a}. This nonlinear behavior can originate from an intrinsic nonlinearity in the cavity design, but is also a fundamental effect of the optomechanical coupling scheme \cite{diaz-naufal}. Thus, by increasing the driving strength, optomechanical setups inevitably enter a regime where the nonlinearity becomes relevant. For backaction cooling, systems residing in the linear regime are commonly considered to be ideal, since they allow for high photon numbers and predictable cooling with negligible nonlinear properties. Thus, optomechanical experiments are typically designed to be as linear as possible and drive strengths are limited to avoid entering the nonlinear regime \cite{Teufel2008DynBA,Teufel_ultrastrong2019}.

In recent measurements \cite{zoepflKerrEnhances2023} we showed that the intrinsic nonlinearity of the cavity can be harnessed to increase cooling capabilities in the bad cavity regime ($ \kappa > \omega_m$) \cite{diaz-naufal,laflammeQuantumlimitedAmplificationNonlinear2011,nationQuantumAnalysisNonlinear2008b}. In this work, we deliberately increase the drive strength above the critical input power of the cavity. There the system undergoes a bifurcation from a monostable to a bistable regime, where the cavity resides either in the low or high photon number branch within a specific range of frequencies. We report on the observation of backaction cooling when driving beyond the critical input power of the cavity as analyzed and predicted in \cite{nationQuantumAnalysisNonlinear2008b}. The theory developed by Diaz-Naufal et al. \cite{diaz-naufal} used within this work describes the observed backaction in the bistable regime depending on the photon state of the cavity.\\

The Hamiltonian for a nonlinear cavity dispersively coupled to a mechanical resonator in the rotating frame of the drive can be written as \cite{diaz-naufal,nationQuantumAnalysisNonlinear2008b}
\begin{equation}
  \hat{\mathcal{H}}/\hbar = -\Delta \hat{a}^\dagger \hat{a} - \frac{\mathcal{K}}{2} \hat{a}^\dagger \hat{a}^\dagger \hat{a}  \hat{a} + \omega_m \hat{b}^\dagger \hat{b} +g_0 \hat{a}^\dagger \hat{a} (\hat{b}+\hat{b}^\dagger) + \hat{\mathcal{H}}_\mathrm{d}
  \label{eq:Ham}
\end{equation} 
with $\Delta = \omega_d-\omega_c$ the drive ($\omega_d$) - cavity ($\omega_c$) detuning, $\omega_m$ the mechanical resonance frequency, $\hat{a}\,(\hat{b})$ and $\hat{a}^\dagger\,(\hat{b}^\dagger)$ the annihilation and creation operators of the cavity (mechanics). The Kerr constant denoted by $\mathcal{K}$ represents the lowest-order nonlinearity of the cavity causing a frequency shift dependent on the photon number. Here, $g_0$ is the single-photon optomechanical coupling strength and $\hat{\mathcal{H}}_\mathrm{d} = \alpha_\mathrm{d} \hat{a}^\dagger + h.c.$ represents the time-independent external drive. Utilizing standard \textit{input-output} theory, the equation of motion of the cavity can be derived as \cite{gardinerInputOutputDamped1985a}
\begin{equation}
  \frac{d}{dt}\hat{a} = -i [\hat{a},\hat{\mathcal{H}}] - \frac{\kappa}{2} \hat{a} -\sqrt{\kappa_c} \hat{a}_\mathrm{in}.
  \label{eq:EOM}
\end{equation}
Here, the total linewidth $\kappa=\kappa_c +\kappa_i$ is introduced as the sum of an intrinsic loss rate $\kappa_i$ and a coupling rate $\kappa_c$ to an input photon field $\hat{a}_\mathrm{in}$ resulting in an input photon number $n_\mathrm{in}$.

\begin{figure}[t]
  \centering
  \includegraphics[width=\columnwidth]{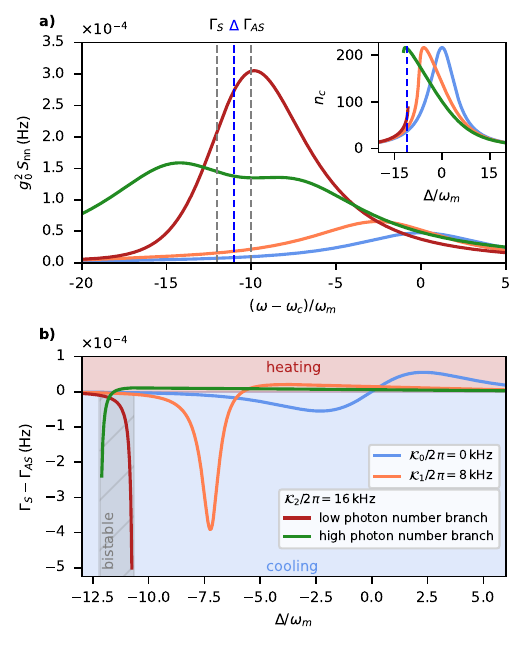}
  \caption{a) Photon number spectrum $g_0^2 S_\mathrm{nn}[\omega]$ for cavities with $\mathcal{K}_0/2\pi = 0 \,\mathrm{kHz}$, $\mathcal{K}_1/2\pi = 8\,\mathrm{kHz}$ and $\mathcal{K}_2/2\pi = 16\,\mathrm{kHz}$. Other parameters for the plots are linewidth $\kappa/2\pi = 3\,\mathrm{MHz}$, mechanical frequency $\omega_\mathrm{m}/2\pi = 300\,\mathrm{kHz}$ and a coupling strength of $g_0/2\pi = 1.7 \,\mathrm{kHz}$. The drive is detuned by $\Delta/\omega_m = - 11$ from the bare frequency of the cavity $\omega_c$ and set to a constant power $n_{in}/n_{bi}=1.5$ with respect to $\mathcal{K}_2$. Gray dashed lines indicate the (anti-)/Stokes scattering processes. The photon induced frequency shift due to $\mathcal{K}$ is evident and becomes more pronounced with increasing $\mathcal{K}$. For $\mathcal{K}_2$, $S_\mathrm{nn}[\omega]$ shows two stable solutions indicating the bifurcation due to the splitting of the photon number $n_c$ shown in the inset. Same color scheme as in b). b) Imbalance of the Stoke and anti-Stokes rates for different detunings $\Delta/\omega_m$. The larger imbalance shows the enhanced cooling capability for a nonlinear cavity in the mono and bistable regime compared to a linear system. For $\mathcal{K}_2$ both solutions feature a discontinuity close to switching to the other branch.}
  \label{fig:scheme}
\end{figure}

Backaction cooling in the unresolved regime can be explained via the conventional scattering framework taking into account the shape of the photon number spectrum $S_\mathrm{nn}[\omega] = \int_{-\infty}^{\infty}\,dt\,e^{i\omega t}\,\langle(\hat{a}^\dagger\hat{a})(t)(\hat{a}^\dagger\hat{a})(0)\rangle$ within the rotating frame of the drive \cite{diaz-naufal,marquardtQuantumTheoryOptomechanical2008}. Due to the optomechanical coupling, photons of an input drive within the cavity resonance are scattered to higher/lower frequencies by annihilation/creation of a phonon in the resonator. These (anti-)Stokes processes occur at a rate $(\Gamma_{AS})\,\Gamma_S$ and are directly proportional to $g_0^2 S_\mathrm{nn}[\omega]$ of the cavity at $\omega = \omega_d \pm\omega_m$ as indicated in Fig. \ref{fig:scheme}a). Here, the spectra $S_\mathrm{nn}[\omega]$ of a linear ($\mathcal{K}=0$) and two nonlinear ($\mathcal{K}>0$) systems are depicted. For a drive detuned by $\Delta/\omega_m  = -11$, $S_\mathrm{nn}[\omega]$ of the linear system has a Lorentzian shape and is symmetric around the resonance of an undriven system. In contrast, for a nonlinear system ($\mathcal{K}>0$) in presence of a strong drive $S_\mathrm{nn}[\omega]$ becomes asymmetric and the resonance frequency shifts. The power-dependent shift is a consequence of the Kerr term in Eq. (\ref{eq:Ham}) influencing $S_\mathrm{nn}[\omega]$. If the drive exceeds the critical input power given by 
\begin{equation}
  n_{bi} = \frac{\kappa}{\kappa_\mathrm{c}}\frac{\kappa^2}{3\sqrt{3}\mathcal{K}},
  \label{eq:n_crit}
\end{equation}
the cavity photon number $n_c$ and thus $S_\mathrm{nn}[\omega]$, split into the low and high photon number branch within a certain frequency range corresponding to the red and green curves in Fig. \ref{fig:scheme}. At these high driving strengths, $S_\mathrm{nn}[\omega]$ deviates from a purely Lorentzian shape as it is visible for the red/green traces in Fig. \ref{fig:scheme}a) \cite{diaz-naufal}. Applying an off-resonance drive tone to the cavity, $S_\mathrm{nn}[\omega]$ around the drive tone leads to an imbalance in $(\Gamma_{AS})\,\Gamma_S$, which results in a net cooling/heating of the phonon mode. In Fig. \ref{fig:scheme}b) this imbalance of $\Gamma_{S}-\Gamma_{AS}$ is tracked for different detunings $\Delta$ of the drive tone with respect to the cavity resonance, where cooling occurs for $\Gamma_{AS}>\Gamma_S$.

For a linear cavity, for red detunings $(\Delta<0)$ cooling and for blue detunings $(\Delta>0)$ heating is observed. Since for a linear cavity $S_\mathrm{nn}[\omega]$ is symmetric around the resonance frequency, the imbalance and the corresponding backaction on the mechanics are also symmetric. In a nonlinear but otherwise identical system, the imbalance $\Gamma_S-\Gamma_{AS}$ is altered due to the asymmetric shape of $S_\mathrm{nn}[\omega]$. This leads to an increase in the cooling capability as illustrated in \cite{diaz-naufal,zoepflKerrEnhances2023,nationQuantumAnalysisNonlinear2008b} and shown in Fig. \ref{fig:scheme}b). The improved cooling capabilities coincide with strong intra-cavity squeezing. However, with a single pump close to resonance we cannot adjust the phase of the squeezing angle independently of the cavity cooling. This means that we cannot utilize the squeezing to counteract backaction heating, but we cool more efficiently due to the enhanced fluctuations of the squeezed cavity \cite{diaz-naufal,asjadOptomechanicalCoolingIntracavity2019,Clerkgroundstate_badcavity2020}. Within the bistable regime of such a system, an abrupt change in the rates can be observed right at the detuning where the photon number jumps from one branch to the other. A drive close to these detunings leads to a suppression of $\Gamma_{S}$ compared to $\Gamma_{AS}$. Thus, in Fig. \ref{fig:scheme}b) a rapid increase in the cooling capability within the proximity of the switching is obtained, similar to the results discussed in \textcolor{red}{\cite{nationQuantumAnalysisNonlinear2008b}}. However, to observe backaction in the bistable regime, it is crucial to be sufficiently close to these detunings. This makes cooling beyond bifurcation of the cavity susceptible to frequency fluctuations, which can cause the cavity to switch branches nullifying the observed backaction. Therefore, it is crucial to establish an experimental setup well isolated from external noise sources, such as flux or vibrational noise, to avoid those switching events.\\

\begin{figure}[t]
  \centering
  \includegraphics[width=\columnwidth]{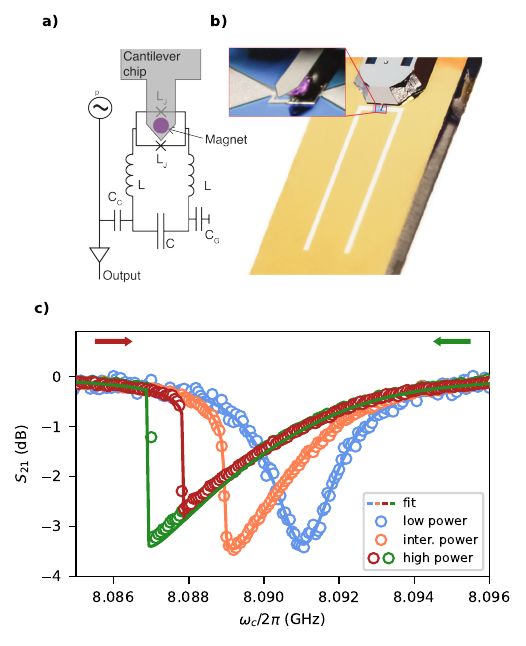}
  \caption{Setup and characterisation. a) Schematic circuit diagram of the setup with junction inductance $L_J$, cavity inductance $L$ and $C$, $C_c$ and $C_g$ as the self, coupling and ground capacitance, respectively.  b) Picture of the sample. Silicon substrate in gold, \textit{U}-shaped cavity in silver, cantilever in gray and magnet false colored in purple (a,b taken from \cite{zoepflKerrEnhances2023}). c) Cavity response for different input powers. For intermediate powers the effects of the nonlinearity become visible and for highest power the cavity shows the splitting into low (red) and high (green) photon number branch. For this power the arrows indicate the direction of the frequency sweep.}
  %c) Resonance frequency change of the cavity by applying an external magnetic field.
  \label{fig:setup}
\end{figure}
The experiment features a flipchip design of a microstrip $\lambda/2$-cavity \cite{zoepflCharacterizationLowLoss2017} coupled to a cantilever as shown in Fig. \ref{fig:setup}a) and b), similar to the setup discussed in \cite{zoepflSinglePhotonCoolingMicrowave2020a,zoepflKerrEnhances2023}. To realize an inductive coupling between cavity and mechanical resonator, a superconducting quantum interference device (SQUID) is embedded in the center of the cavity making it flux sensitive, similar to other systems based on an inductive coupling scheme \cite{bothnerFourwavecoolingSinglePhonon2022,luschmannMechanicalFrequencyControl2022,beraLargeFluxmediatedCoupling2021}. By attaching a magnetic particle to the tip of the cantilever its motion modulates the magnetic flux through the SQUID, which couples the two oscillator modes. The chip is placed inside a 3D-waveguide with a central coil around it, allowing us to apply an external DC magnetic field to insitu tune the cavity frequency and thus the optomechanical coupling strength from a few $\mathrm{Hz}$ up to $g_0 \approx 90\,\mathrm{kHz}$ \cite{zoepflSinglePhotonCoolingMicrowave2020a,zoepflKerrEnhances2023}. The whole setup is mounted to the base of a dilutional cryostat operated at $100\,\mathrm{mK}$. Additionally, to decouple our system from any mechanical vibrations we implemented a vibration isolation setup consisting of a long nylon wire and a stainless steel spring acting as a low pass filter with a resonance frequency of $\omega_z/2\pi \approx 2.4\,\mathrm{Hz}$ (see Appendix \ref{sec:Setup}). The cavity has a frequency of $\omega_c/2\pi = 8.1\, \mathrm{GHz}$, and a total linewidth of $\kappa/2\pi = 2.8\,\mathrm{MHz}$, whereas the cantilever has a frequency of $\omega_m/2\pi = 287.3\,\mathrm{kHz}$ and a linewidth of $\Gamma_m/2\pi = 0.4\,\mathrm{Hz}$. 

The nonlinear behavior of the cavity can be visualized by scanning its response for different probe powers as shown in Fig. \ref{fig:setup}c). By increasing the power we first observe a frequency shift of the cavity response accompanied by an asymmetric shape of the resonance. Further increasing the drive power above $n_{in,bi}$ one can see the splitting of the cavity response into the two branches. For both branches the detuning, where the cavity switches into the other branch, is visible by a discontinuous jump in the response. Depending on the direction of the frequency sweep, represented by the arrows in Fig. \ref{fig:setup}c), the different states can be accessed.
\begin{figure}[b]
    \centering
    \includegraphics[width=\columnwidth]{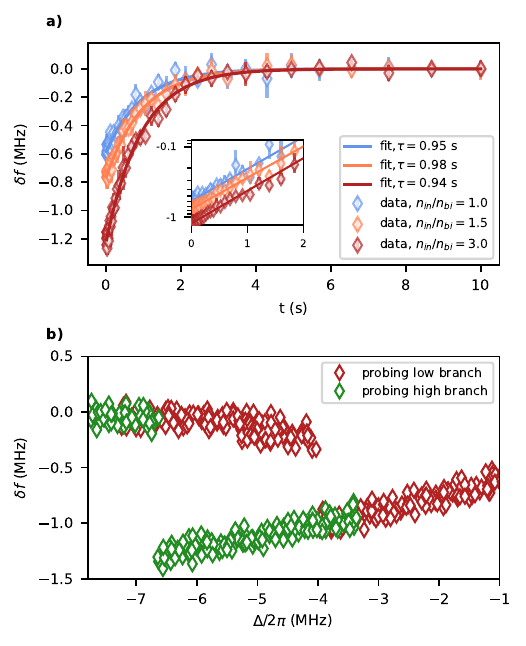}
    \caption{a) Relaxation of the frequency shift $\delta\mspace{-4mu}f$ after driving the cavity at different powers, for the cavity being initialised in the high photon number branch. Solid lines represent an exponential fit to extract the timescale $\overline{\tau} = (0.96\pm 0.02) \,\mathrm{s}$. Inset shows a zoom in on the data plotted on a log-scale. b) Frequency shift $\delta\mspace{-4mu}f$ against probe-cavity detuning $\Delta$ for a drive strength $n_{in}/n_{bi} = 3.0$. The differen branches are probed by tuning the probe towards the cavity from low/high frequencies to scan the low (red) and high (green) branch.}
    \label{fig:shift}
\end{figure}
Additionally to the Kerr-induced frequency shift shown in Fig. \ref{fig:setup}d), we observed another frequency shift $\delta\mspace{-4mu}f$ of the cavity. This shift can be measured by weakly probing the cavity response in presence of a strong drive tone. By populating the cavity, the resonance frequency shifts until it reaches a new steady state on the timescale of seconds. With higher drive strengths or placing the drive closer to the resonance, both effectively increasing the intracavity photon number, $\delta\mspace{-4mu}f$ between the new steady state $f_{ss}$ and initial $f_{i}$ frequency grows. To quantify the timescale of this shift, we measure the evolution of the cavity response after turning off the drive tone. As shown in Fig. \ref{fig:shift}a) $\delta\mspace{-4mu}f(t)$ follows an exponential decay back to its initial value. A fit to the data yields a slow relaxation time of $\overline{\tau} = (0.96\pm 0.02) \,\mathrm{s}$ independent of the photon number, whereas the amplitude of $\delta\mspace{-4mu}f$ significantly depends on it.

We can use this photon number dependence of the frequency shift to determine in what state the cavity ended up after applying a drive. Since the photon number in the low and high photon number branch significantly differs, $\delta\mspace{-4mu}f$ measured right after turning off the drive will be large in the high and small for the low photon number branch. In Fig. \ref{fig:shift}b) this is shown by displaying $\delta\mspace{-4mu}f = f_{ss}-f_{i}$ against cavity-drive detuning $\Delta$ for a drive exceeding the critical input power. While residing in the low photon number branch, the cavity is almost not populated resulting in no detectable shift. However, close to switching to the high photon number branch a small shift is observed followed by a distinct frequency jump upon switching. Scanning the high state, we observe a large frequency shift due to the highly populated cavity until a detuning is reached, where the cavity switches into the low photon number branch visible by $\delta\mspace{-4mu}f \approx 0$. Between the switching events of the branches, the cavity shows two distinct shifts associated with either one of the cavity states. Although the origin of the observed shift remains unclear (see Appendix), we can take it into account by adjusting the measurement procedure to contain wait times after turning on the drive to ensure a steady state of our cavity.\\

To characterize the mechanics, we measure the homodyne noise spectrum of a probe tone applied to the cavity. Due to the coupling of the cavity to the mechanics, the probe tone undergoes an effective amplitude (phase) modulation at the mechanical frequency $\omega_m$ passing through the system \cite{zoepflSinglePhotonCoolingMicrowave2020a}. To extract the phonon occupation $\langle n_m\rangle$, mechanical frequency $\omega_m$, and the linewidth $\Gamma_m$ we fit the model of a damped harmonic oscillator to the resulting sidebands and calibrate the signal via a method derived in \cite{gorodetksyDeterminationVacuumOptomechanical2010a}. By measuring the mechanical signature on the probe tone for different drive-cavity detunings $\Delta = \omega_d - \omega_c$ the backaction on the mechanics can be studied by tracking $\langle n_m\rangle$, $\omega_m$ and $\Gamma_m$. To determine the influence of the intrinsic cavity nonlinearity we take cooling traces for different input powers. Furthermore, to ensure a steady state of the cavity we slowly tune the drive into resonance in the same directions as shown in Fig. \ref{fig:setup}c) to select the low/high photon number branch beyond bifurcation. Additionally, we utilize the discussed properties of $\delta\mspace{-4mu}f$ to determine the cavity state during the measurement to avoid data points containing a switching event.\\

%Because of its small linewidth of $\Gamma_m/2\pi = 0.4\,\mathrm{Hz}$ the acquisition of a single mechanical spectrum takes $\approx 10\,\mathrm{s}$, switching events can occure while taking the mechanical spectrum. By measuring the cavity response right after taking the mechanical spectrum, we can utilize the discussed properties of $\delta\mspace{-4mu}f$ to determine the cavity state during the measurement to avoid data points containing such a switching event.\\
%\onecolumngrid
\begin{figure*}[t!]
  \centering
  \includegraphics[width=2\columnwidth]{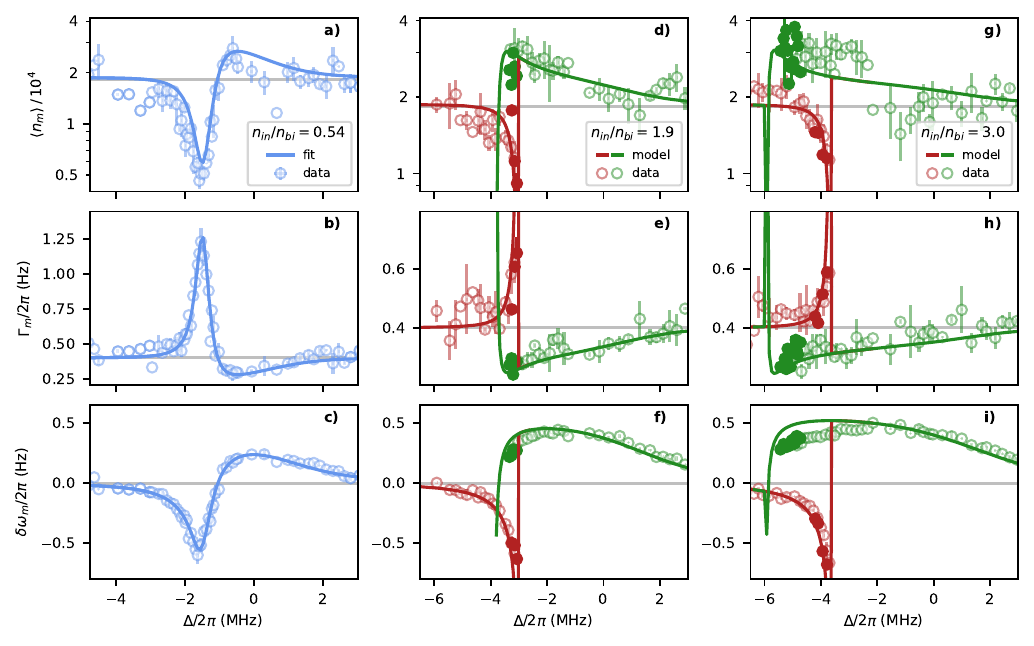}
  \caption{Cooling traces at $g_0/2\pi = 99\,\mathrm{Hz}$. a) Phonon number $\langle n_m \rangle$, b) linewidth $\Gamma_m/2\pi$ and c) frequency shift $\delta \omega_m /2\pi$ for intermediate powers $n_{in}/n_{bi}=0.5$. Gray lines display a steady state at an effective bath temperature $T_\mathrm{eff} = 267\,\mathrm{mK}$. Phonon number $\langle n_m \rangle$, linewidth $\Gamma_m/2\pi$ and frequency shift $\delta \omega_m /2\pi$ for higher powers of $n_{in}/n_{bi}=1.9$ in  d-f) and $n_{in}/n_{bi}=3.0$ in g-i), respectively. Data and theoretical extrapolation from $n_{in}/n_{bi}=0.5$ in red for the low and green for the high photon number branch. Filled circles correspond to data taken with the pulse tube of the cryostat turned off to further reduce vibrations.}
  \label{fig:by_cooling}
\end{figure*}
%\twocolumngrid

%For comparison a theoretical cooling trace of a linear but otherwise identical system is displayed, visualizing the effect of the nonlinearity on the backaction.
Figures \ref{fig:by_cooling}a-c) show a cooling trace at moderate coupling strength of $g_0/2\pi = (99 \pm 1)\,\mathrm{Hz}$ to minimize flux noise effects. Fitting the data with a theory derived in \cite{diaz-naufal} we observe good agreement with the experimental results. The fit yields an input photon number of $n_{in}/n_{bi}=(0.54\pm0.02)$ and a Kerr constant of $\mathcal{K}/2\pi = (14\pm1)\,\mathrm{kHz}$. Already at these drive powers, the cooling trace shows the effects of the nonlinearity, namely an asymmetric backaction trace with a narrow cooling window. The displayed thermal phonon number (gray line) corresponds to an extracted effective bath temperature of $T_\mathrm{eff} = 267\,\mathrm{mK}$ even though we set the base temperature of the fridge to $100\,\mathrm{mK}$ (see. Appendix \ref{sec:tempramp}).

Cooling traces for increased input powers corresponding to $n_{in}/n_{bi} = 1.9$ and $n_{in}/n_{bi} = 3.0$ are shown in Fig. \ref{fig:by_cooling}d-f) and i-g), respectively. For the different input powers, over a detuning range of about $0.4\,\mathrm{MHz}$ and about $ 2\,\mathrm{MHz}$ we observe two distinct backaction curves of the mechanics depending on the selected branch. The solid lines in the plots depict the theoretical prediction extrapolated from the parameters obtained from the fit in Fig. \ref{fig:by_cooling}a-c). Even though, we increased the input power by a factor of about $4$ and about $6$, resulting in a cavity well beyond the bifurcation point, the prediction accurately describes the data for the two different branches. In both cases, the backaction in the low photon number branch exhibits a narrow cooling feature up to the point at which the cavity switches to the high photon number branch having reached the end of the bifurcation region. Upon switching, the cavity resides in the high photon number branch, where the photon number spectrum of the cavity results in backaction heating. The experimental data obtained agree well with the predictions and the detuning at which branch switching is observed matches with the theory.

In case of the high photon number branch we observe heating over a large frequency range before the cavity state switches. The high photon number branch shows a very narrow cooling feature that cannot be experimentally accessed. This is due to constant heating of the mechanical mode, leading to a highly excited state as we tune the probe tone slowly in from the blue side to ensure staying in the high photon number branch. The corresponding large motional amplitude leads to fluctuations of $\omega_c$, which results in the cavity switching branches before reaching the cooling feature in the high photon number branch. The experimental results obtained for the low and high photon number branch demonstrate that we have reliable control over the cavity state in the bistable regime except for a narrow frequency range close to the bistable point, where residual noise in the system causes random switching between the branches. Furthermore, the theory predictions match with the data for all measured input powers (see Appendix \ref{sec:all_traces}), which underlines a good understanding of the system studied.

Summarizing, we demonstrated that we can operate our system even beyond the bifurcation of the cavity and expand the observation on backaction cooling for nonlinear systems made in \cite{zoepflKerrEnhances2023}. Due to the control over the selected branches in this regime we achieved good agreement with the underlying theory. By extrapolating fit results over a factor of $6$ in input power we can precisely predict the cooling behavior obtained, demonstrating a good understanding of processes at hand. Furthermore, the observed cooling behavior beyond bifurcation originates due to a deviation of the cavity photon number spectrum from the typical Lorentzian shape. An interesting fact is that the best cooling is theoretically and experimentally obtained in the low instead of the high photon number branch, contrary to the common opinion that an increased photon number should lead to an enhanced cooling.  Overall, the resulting cooling in the bistable regime shown here is limited by flux noise in the system, preventing us from precisely accessing the narrow frequency range in which best cooling is expected. Also, further investigations have to be made to grasp the origin of the observed slow frequency shift of our cavity. Nevertheless, the presented results demonstrate that state manipulation in optomechanical systems should not be restricted only to the linear regime. Exploiting the nonlinearities may offer novel mechanisms to be explored for optomechanical systems in the unresolved sideband regime.

\begin{acknowledgments}
We want to thank Hans Huebl and Christian Schneider for the fruitful discussions. Further we thank our in-house workshop. This research was funded in whole or in part by the Austrian Science Fund (FWF) [Grant DOI:10.55776/W1259]. For open access puposes the author (L.D.) has applied a CC BY public copyright licence to any author-accepted manuscript version arising from this submission. N.D.N and A.M. acknowledge funding by the project CRC 183. A.M. acknowledges funding by the Deutsche Forschungsgemeinschaft through the Emmy Noether program (Grant No. ME4863/1-1). M. L. J. acknowledges funding by the Canada First Research Excellence Fund. This work was supported by the European Union‘s Horizon Europe 2021-2027 Framework Programme under Grant Agreement No. 101080143 (SuperMeQ).
\end{acknowledgments}
The data that support the plots within this paper and the appendix will be available via a permanent, public repository, Zenodo DOI 10.5281/zenodo.11502595.
%It remains to show, if the combination of a nonlinear system with squeezed light allows to reach the groundstate in the unresolved regime as postulated in \textbf{Nico}. As cancelation of backaction in a linear system by using externally squeezed light could be shown in \textbf{ref}, it would be interesting to compare this with a system containing a nonlinear cavity. Further, investigating effects on our cooling capabilites of this squeezing with internal generated squeezing enabled by the Kerr nonlinearity would be interesting.  
%%%%%%%%%%%%%%%%%%%%%%% experiment

\bibliography{cited_references}

\newpage

\onecolumngrid
\appendix
%\title{Appendix: Optomechanical Backaction in the Bistable Regime}

%\maketitle
    
%\onecolumn
\setcounter{section}{0}
\setcounter{equation}{0}
\renewcommand{\thesection}{S. \arabic{section}}
\renewcommand{\theequation}{S. \arabic{equation}}
\setcounter{figure}{0}
\renewcommand{\thefigure}{S. \arabic{figure}}

\section{Setup}\label{sec:Setup}

\begin{figure}[h!]
    \centering
    \includegraphics[width=\columnwidth]{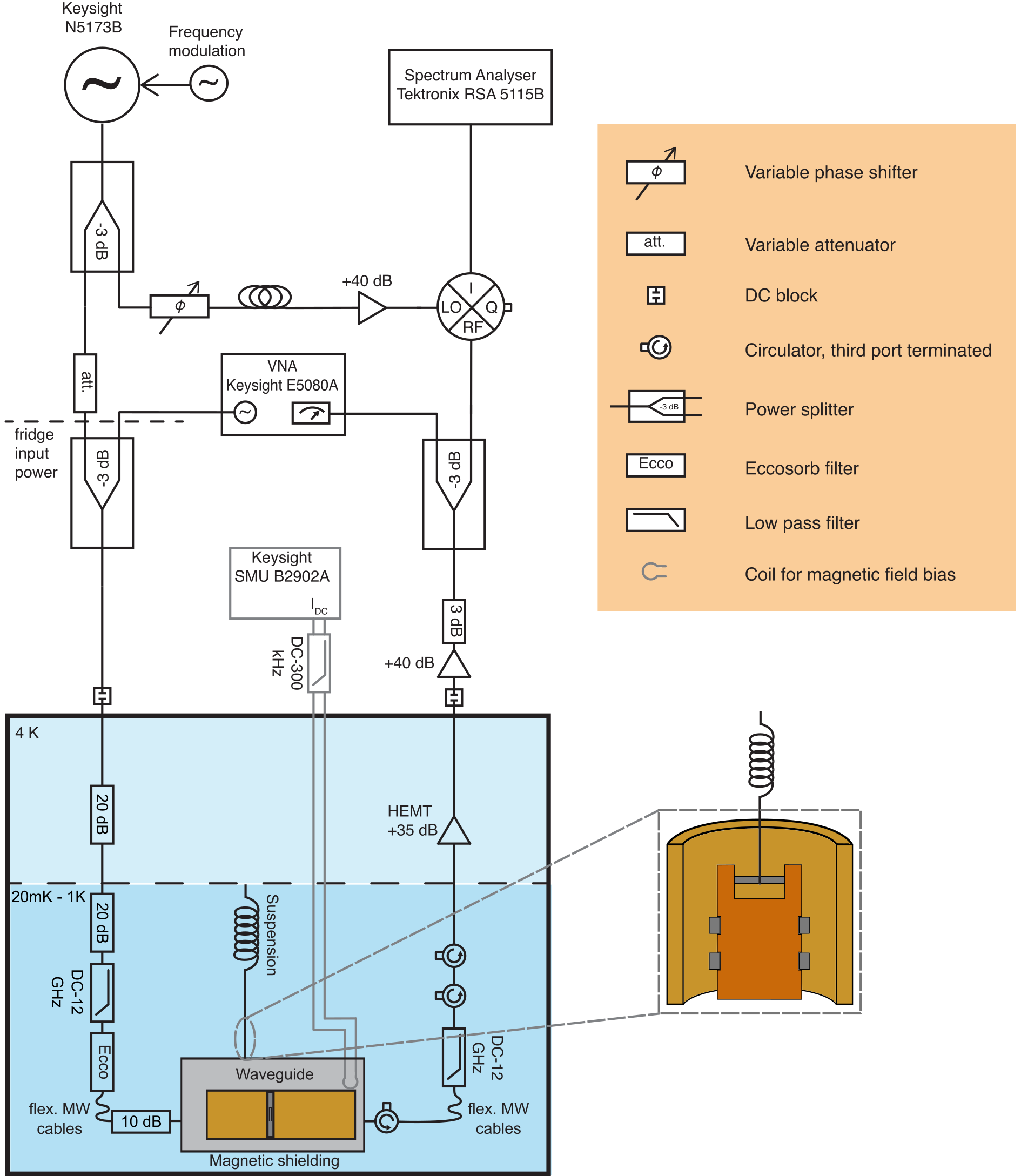}
    \caption{Schematic drawing of the setup. The dashed circle indicates the positions of the eddy current damping tube not shown in Fig. \ref{fig:susp_fridge}a). The spring is connected to the setup via a tower structure inside a brass tube attached to the base plate of the cryostat. For the damping we then attach several NdFeB magnets (grey blocks in the zoom in).}
    \label{fig:setup_fridge}
\end{figure}

\begin{figure}[h!]
    \centering
    \includegraphics[width=\columnwidth]{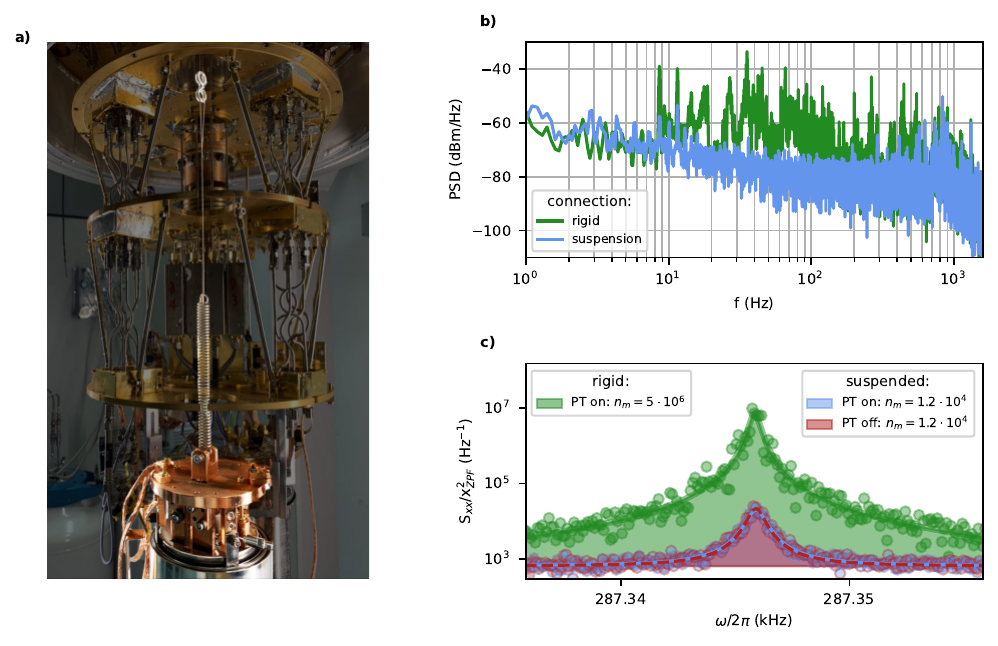}
    \caption{a) Photograph of the suspended setup with the spring and the wire highlighted. At the bottom the beginning of the magnetic shield is visible. b) Room temperature accelerometer measurements to compare the vibrational noise for a system rigidly connected to the base plate of the cryostat and one suspended from the still plate. Above $1\,\mathrm{kHz}$ the detection is limited by the cable connected to the accelerometer. c) Mechanical spectrum of the resonator at $T_\mathrm{bath} = 150\,\mathrm{mK}$ measured with running pulsetube and pulsetube turned off for a rigidly connected and suspended setup.}
    \label{fig:susp_fridge}
\end{figure}

The sample as well as the measurement setup (Fig. \ref{fig:setup} and Fig. \ref{fig:setup_fridge}) are very similar to the one described in \cite{zoepflSinglePhotonCoolingMicrowave2020a,zoepflKerrEnhances2023}. To decouple our system from vibrations of the fridge we implemented a one-stage suspension setup consisting of a nylon wire and a steel spring. To achieve reasonable damping in the \textit{xyz}-plane the setup is suspended from the still plate as shown in Fig. \ref{fig:susp_fridge}a). To keep the vibrations coupled in over the input/output lines low we used cryoflex cables from \textit{Delft Circuits} and to ensure thermalization of the setup we used 3 flexible oxygen-free high-conductivity copper braids.\\
Previously, we suffered from vibrational noise at $1\,\mathrm{Hz}\,-\,1\,\mathrm{kHz}$ \cite{zoepflKerrEnhances2023} when the pulse tube cooler was operating. The suspension setup significantly decouples our system in this frequency range as shown in Fig. \ref{fig:susp_fridge}b). To characterize the isolation we briefly turned on the pulse tube cooler at room temperature and measured the mechanical vibrations with an accelerometer attached to the setup, either rigidly connected to the base or suspended from the still plate. With the suspension setup, we achieve a damping of the vibrational noise in the low frequency range up to $1\,\mathrm{kHz}$ by $\sim20\,\mathrm{dBm}$. Above $1\,\mathrm{kHz}$ the detection is limited by the cable connected to the accelerometer. Furthermore, to prevent a ring up scenario of the suspended setup at its resonant frequency, we implemented an eddy current damping tube [not shown in Fig. \ref{fig:susp_fridge}a) as it was implemented later]. Fig. \ref{fig:susp_fridge}c) depicts a mechanical spectrum of the cantilever at $150\,\mathrm{mK}$ measured with the pulse tube turned on and off. Since both spectra match and the mechanical occupation obtained in both cases is close to the thermal occupation of $n^\mathrm{th}_\mathrm{m} = 1.1\cdot10^4$, the suspension setup allows us to measure the mechanical signal with the pulse tube running, which was previously not possible. This leads to an increase in the data acquisition rate as well as allowing us to measure the data for the backaction beyond bistability due to the increased stability.

\section{Data evaluation and Treatment}\label{sec:DT}
\subsection{Cavity}
The cavity parameters can be extracted using the circle fit routine for a response measured with a vector network analyzer (VNA). The transmission $ S_{21}$ in a hanger-notch-configuration is then given by \cite{probstEfficientRobustAnalysis2015}
\begin{equation}
    S_{21}(\omega) = a e^{i\alpha} e^{-i\tau \omega} \left( 1 - \frac{Q_l/|Q_c|e^{i\phi_0}}{1+2iQ_l\frac{\omega-\omega_c}{\omega}}\right)
    \label{eq:circle}
\end{equation}
Within the parenthesis is the ideal response of a resonator in transmission with the loaded $Q_l$ and coupled quality factor $Q_c$, the impedance missmatch $\phi_0$ and the resonance frequency $\omega_c$. Environmental effects are taken into account by $a$ and $\alpha$, while $\tau$ quantifies the electric delay.\\
\begin{figure}[h!]
    \centering
    \includegraphics[width=\columnwidth]{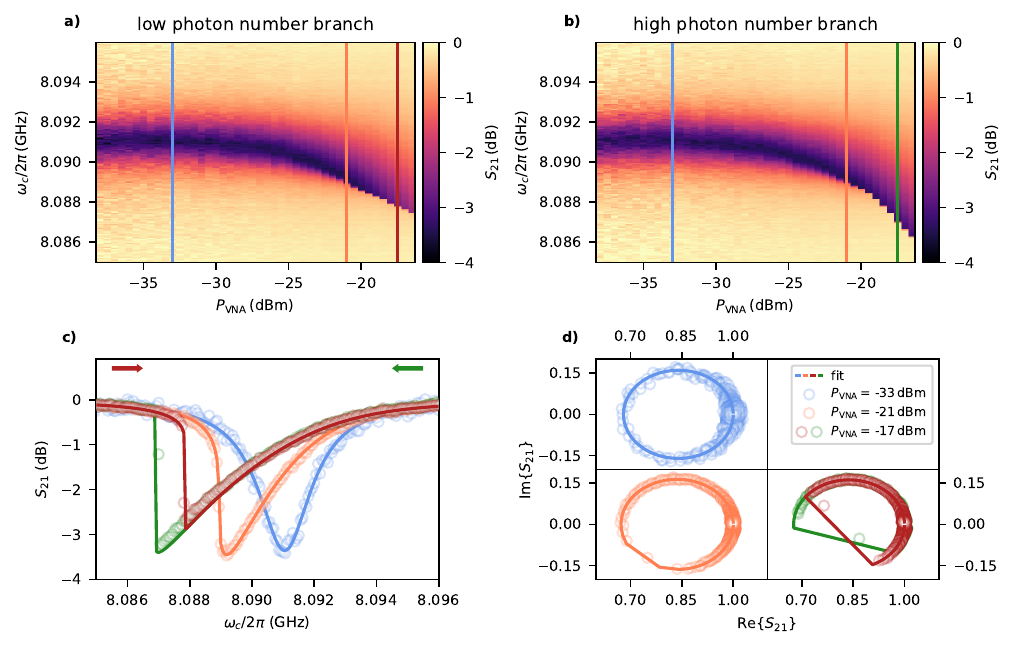}
    \caption{a) and b) Cavity response with increasing drive strength scanning the low and high branch, respectively. c) Amplitude data for 3 different powers, where for highest powers the signal bifurcates. Low and high branch accessed by scanning the drive from different directions indicated by the arrows. d) Data in the complex plane.}
    \label{fig:circle}
\end{figure}
A good estimate for the Kerr constant can be achieved by adapting the circle fit routine by adding the photon number induced cavity shift of a nonlinear cavity. For that the resonance frequency $\omega_c$ in Eq. (\ref{eq:circle}) has to be transformed to 
\begin{equation}
    \omega_c \to \,\omega_c - \mathcal{K} n_c
    \label{eq:kerrshift}
\end{equation}
\begin{equation}
    n_c[(\omega_c - \mathcal{K} n_c -\omega)^2+ (\frac{\omega_c}{2 Q_l})^2] = \frac{\omega_c}{2|Q_c|} \frac{P_g}{\hbar \omega}        
    \label{eq:nonlincircle}
\end{equation}
with the Kerr constant $\mathcal{K}$, the intracavity photon number $n_c$ and the drive power $P_g$ reaching the cavity. The intracavity photon number for a nonlinear system is given by the cubic equation in Eq. \ref{eq:nonlincircle} \cite{diaz-naufal}. To be compatible with the input-output theory convention leading to the $S_{21}$ response given in Eq. (\ref{eq:circle}) a factor $1/2$ has to be added for the photon input. By taking the input attenuation into account one can extract the Kerr constant by fitting consecutive cavity responses with increasing power Fig. \ref{fig:circle}. The fit yields a Kerr constant of $\mathcal{K} = (12 \pm 4)\,\mathrm{KHz/photon}$ being in good agreement with the results observed by fitting the cooling traces in the main text. The given error results from the uncertainty of the input attenuation, assumed to be $\pm 2\,\mathrm{dB}$. 

\subsection{Mechanics}
The measurements and data evaluation for the mechanical data are not significantly changed from the procedure presented in the Appendix of \cite{zoepflKerrEnhances2023}. The only addition to the measurements was the slow tune in of the drive tone before measuring the mechanical spectra and a fast VNA trace triggered by turning off the drive to determine the frequency shift $\delta\mspace{-4mu}f$. For the data evaluation beyond bistability the spectra were additionally sorted corresponding to the state of the cavity based on the $\delta\mspace{-4mu}f$ before using the binning method.

\section{Temperature scan}\label{sec:tempramp}
\begin{figure}[h!]
    \centering
    \includegraphics[width=\columnwidth]{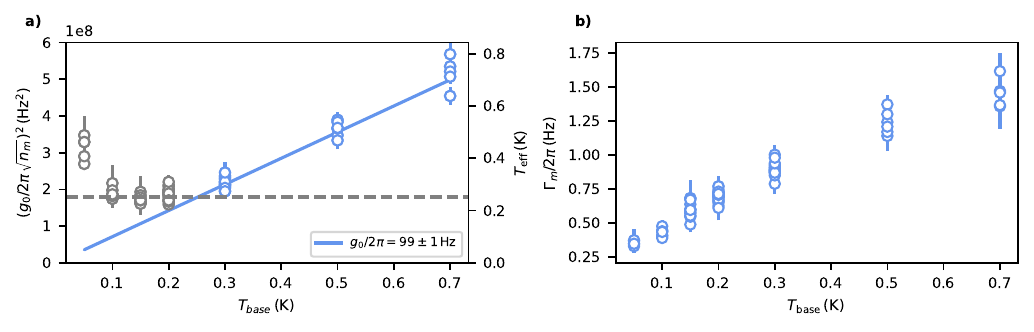}
    \caption{a)Mechanical occupation $g_0^2n_{m}$ measured at different base temperatures $T_\mathrm{base}$ with the cavity bias frequency used in the main text. Colored data points used to calibrate $g_0$. Grey data points indicate, that the mechanical mode does not thermalize to the base temperature. Horizontal line represents the effective bath temperature $T_\mathrm{eff}$ at $T=100\,\mathrm{mK}$. b) The mechanical linewidth against base temperature.}
    \label{fig:temp_ramp}
\end{figure}

Using the heterodyne detection scheme proposed by \cite{gorodetksyDeterminationVacuumOptomechanical2010a} we only have access to $g_0^2 \langle n_{m} \rangle$ by measuring the area below the mechanical signal. Thus, it is essential to know our coupling strength $g_0$ at a given bias frequency to be able to determine $\langle n_{m} \rangle$ for backaction measurements. By assuming that the mechanical mode thermalizes with the base temperature $T$ of the cryostat the phonon occupation is given by the Boltzmann occupation $\langle n^{th}_{m} \rangle \approx \frac{k_B T}{\hbar \omega_m}$. Therefore, measuring $g_0^2 \langle n^{th}_{m} \rangle$ for different temperatures $T$ allows us to extract the single-photon coupling strength $g_0$ at a given bias frequency. To ensure that the mechanical mode is in thermal equilibrium we measure the mode by using a weak probe tone tuned on resonance to avoid backaction. The results are shown in Fig. \ref{fig:temp_ramp}.\\
For temperatures below $250\,\mathrm{mK}$ the data indicates that the mechanical mode does not thermalize to the base temperature, similar to observations for other mechanical systems at low temperature \cite{seis2022ground}. However, for temperatures above $250\,\mathrm{mK}$ $g_0^2 \langle n^{th}_{m} \rangle$ follows a linear trend allowing us to extract $g_0=99\pm1\,\mathrm{Hz}$, being consistent with previous measurements \cite{zoepflSinglePhotonCoolingMicrowave2020a,zoepflKerrEnhances2023} where the cantilever thermalized. Since the coupling strength at this bias frequency should not change with temperature, we can extract an effective bath temperature of the mechanical mode to be $T_\mathrm{eff} = 267\,\mathrm{mK}$ at a base temperature of $T = 100\,\mathrm{mK}$. The observed baseline of the cooling traces for different drive powers agrees with the extracted $T_\mathrm{eff}$, making the assumption of an effectively increased bath temperature of the mechanics reasonable. Furthermore, the observed linewidth follows a temperature dependence similar to previous measurements in \cite{zoepflSinglePhotonCoolingMicrowave2020a,zoepflKerrEnhances2023} with $\Gamma_m/2\pi = 0.4 \, \mathrm{Hz}$ at $T = 100\,\mathrm{mK}$. This further reinforces the assumption of an effective bath temperature acting as an additional phonon reservoir for our resonator.

\section{Frequency shift $\delta\mspace{-4mu}f$}\label{sec:freqshift}

\begin{figure}[h!]
    \centering
    \includegraphics[width=\columnwidth]{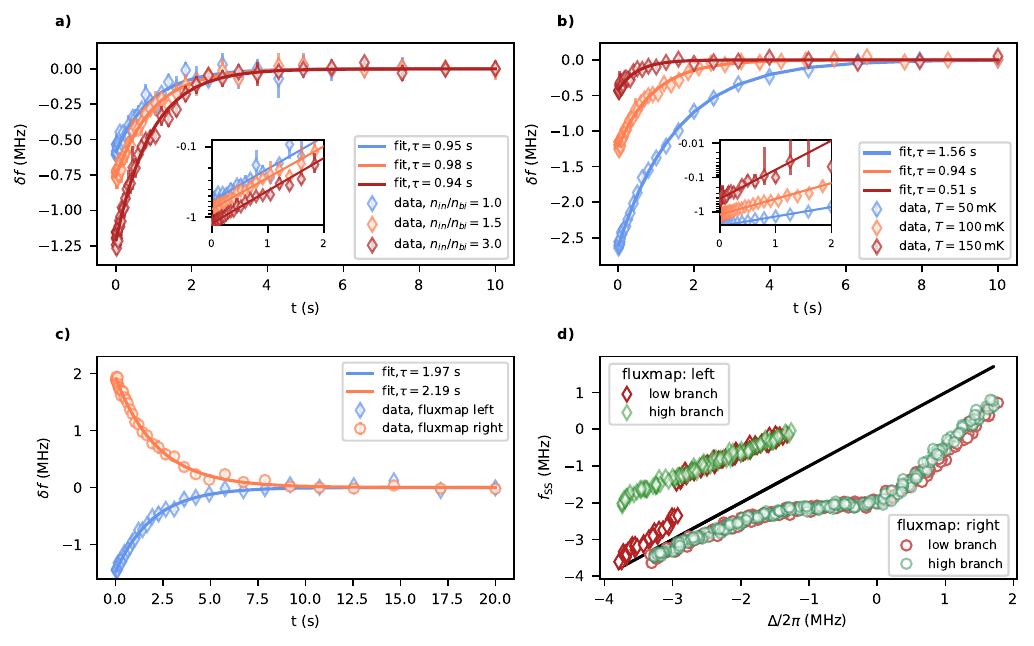}
    \caption{a) relaxation of the frequency shift $\delta\mspace{-4mu}f$ for different drive powers at $T_\mathrm{base}= 100\,\mathrm{mK}$, b) at different base temperature for the same drive power $n_{in}/n_{bi} = 3.0$. c) Comparison of the relaxation process measured on the left side - as in the main text - and right side of the fluxmap at $T_\mathrm{base}= 50\,\mathrm{mK}$ and $n_{in}/n_{bi} \approx 1.0$. d) Reached steady state frequency at the same power as in c) against set detuning $\Delta/2\pi$ while probing the low and high photon number branch for left and right side of the fluxmap. Data above the black line correspond to a positive and below to a negative shift.}
    \label{fig:shift_mem}
\end{figure}
As mentioned in the main text, we observed a frequency shift $\delta\mspace{-4mu}f$ of the cavity on timescales of seconds when strongly driven. Populating the cavity causes the resonance frequency to shift reaching a new steady-state frequency, where the absolute shift increases with cavity population. To quantify the timescale, we measure the relaxation process towards the initial frequency after turning off the applied drive by weakly probing the cavity. In Fig. \ref{fig:shift_mem}a) the dependence of this relaxation process is shown for different drive powers. Since only the absolute shift and not the timescale changes with power, we can use this behavior to distinguish if the cavity resided in the low or high photon number branch during backaction measurements as stated in the main text.\\
Furthermore, the observed frequency shift also prevents continuously switching between the states during a measurement for detunings within the bistable range. Since the cavity shifts toward lower frequencies with increasing intracavity photon numbers, switching from the low to the high photon number branch causes the cavity to shift, effectively increasing the detuning between the drive and frequency associated to the switching event. Changing branch in the opposite direction, the strongly shifted cavity returns to its initial frequency, resulting in a drive far detuned from resonance. Thus, in both cases switching leads to a frequency shift that precludes the restoration of the initial state.\\ 

However, to further investigate the origin of this shift we measured $\delta\mspace{-4mu}f$ for different temperatures as depicted in Fig. \ref{fig:shift_mem}b). Keeping the input power constant at $n_{in}/n_{bi} = 3.0$, we observe that with increasing temperature there is a reduction in the timescale as well as the total frequency shift. Furthermore, when the cavity is tuned to the right side of the flux map instead of the left side, a positive frequency shift is observed (Fig. \ref{fig:shift_mem}c). Taken at the same bias frequency and with the same input power, fitting the relaxation rate yields comparable parameters regardless of the side on which it is measured.\\

Contrary to the behavior on the left flux map side, the positive frequency shift on the right side does not allow the different branches to be distinguished, as shown in Fig. \ref{fig:shift_mem}d). Irrespective of the branch which is probed, the cavity reaches the same steady-state frequency for a broad range of initial frequencies. For high enough input powers $n_{in}/n_{bi} > 1$, a continuous switching between the branches can be observed as the shift causes the drive to tune in and out of the resonance. With this behavior meaningful measurements beyond bifurcation on the right side of the flux map are not possible. However, for input powers $n_{in}/n_{bi} < 1$ the positive shift acts as an effective negative feedback loop stabilizing the resonance against slow frequency fluctuations and thus proves beneficial for backaction measurements before bistability, similar to observations made in \cite{bothnerFourwavecoolingSinglePhonon2022}.\\

Overall, the observed frequency shift scales with the intracavity photon number. Amplitude and timescale decrease for higher temperatures and the direction of the shift depends on the sign of the flux map slope. A frequency shift due to radiation pressure on the mechanical resonator seems plausible at first glance. Frequency tuning of our cavity by applying an external magnetic field leads to a circulating current within the SQUID due to flux quantization. The corresponding magnetic field of the SQUID displaces the cantilever from its resting position. However, increasing the intracavity photon number by driving the cavity decreases the circulating current in the SQUID and the corresponding magnetic field. This in turn leads to a displacement of the mechanical resonator toward its initial resting position. Simultaneously, this changes the constant part of the magnetic flux from the magnetic tip inside the SQUID resulting in a shifted resonance of the cavity. Increasing the cavity population would lead to a lower circulating current, which agrees with the observation, that the amplitude of the frequency shift increases with input power and that the timescale is unaffected by it. Since the linewidth of the mechanical resonator grows significantly with temperature as shown in Fig. \ref{fig:temp_ramp}b), it seems plausible to assume, that this in turn causes the temperature dependence of the timescale on which the shift $\delta\mspace{-4mu}f$ is observed in Fig. \ref{fig:shift_mem}b). Interestingly, the relative changes in $\Gamma_m$ match with the changes in $\tau$, however the absolute values are off by a factor of $\sim 2$. Nevertheless, for the radiation pressure model the frequency shift should scale with $\delta\mspace{-4mu}f \propto -g_0^2$, which contradicts the difference in shift directions depending on the side of the flux map.\\

Another explanation could be that the Kerr constant $\mathcal{K}$ of the cavity is altered by the optomechanical coupling as stated in \cite{diaz-naufal}, given by 
\begin{equation}
    \mathcal{K}_\mathrm{eff} = \mathcal{K}+\frac{2g_0^2\omega_m}{\omega_m^2+\frac{\Gamma_m^2}{4}}.
\end{equation} 
However, the change to $\mathcal{K}$ in the regime we operate the system at would be too small to explain the observed shift and also fails to describe the flux-map-side dependence due to its $g_0^2$ scaling.\\

Also magnetic surface spins are candidates to explain the properties of the frequency shift. At low temperatures the spins would be aligned due to the close proximity of the magnet at the cantilever tip. The ensemble of spins creates a constant offset in magnetic flux at the SQUID loop. However, strongly driving our cavity would cause some of the spins to scramble, which would lead to a decrease in the offset magnetic flux at the SQUID. This would result in a horizontal shift of the flux map towards absolute zero flux (e.g. flux map moves to the left in Fig. \ref{fig:setup}c)). Thus the frequency on the left and right flux map side would shift in $\delta\mspace{-4mu}f<0$ and $\delta\mspace{-4mu}f>0$, respectively, matching the observations. Furthermore, larger input powers and therefore higher alternating currents would create a large magnetic field change, which would cause more spins to be scrambled. Therefore, the offset magnetic field would decrease and thus shift the flux map by a larger amount. This would also match with the obtained power dependence of the frequency shift. Lastly, the relaxation time of the spins should feature the same timescale dependence on temperature as observed for the frequency shift, and, since higher temperatures allow for a higher degree of initial disorder, it would also explain the change in amplitude. Nevertheless, to quantify the influence on surface spins experimentally proves to be challenging and further experimental advances have to be made to verify this hypothesis.

\section{Cooling traces}\label{sec:all_traces}
\begin{figure}[h]
    \centering
    %\vspace{-100pt}
    \includegraphics[height=0.775\paperheight]{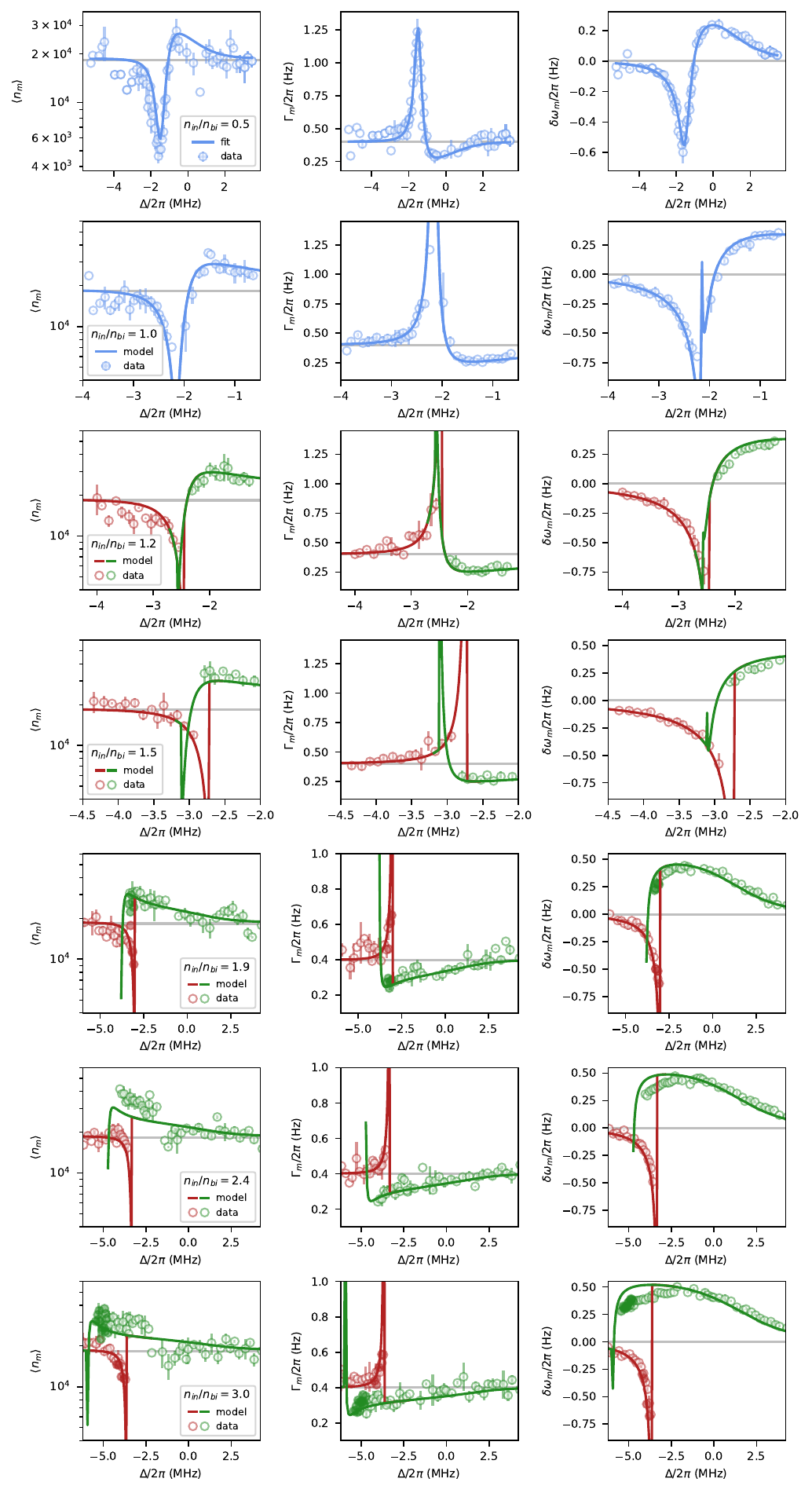}
    \caption{Cooling traces at $g_0/2\pi = 99\,\mathrm{Hz}$ with phonon number $\langle n_m \rangle$, linewidth $\Gamma_m/2\pi$ and frequency shift $\delta \omega_m /2\pi$ for various input powers $n_{in}/n_{bi} = 0.5-3.0$. The cooling trace at $n_{in}/n_{bi} = 0.5$ is fitted to extract the system parameters. For all other input powers the lines depict predictions from the low power fit. Filled circles correspond to data taken with the pulsetube of the cryostat turned off.}
    \label{fig:all_log}
\end{figure}

In Fig. \ref{fig:all_log} additional cooling traces for input powers ranging from $n_{in}/n_{bi} = 0.5$ to $3.0$ are shown, matching with the ones displayed in the main text. The trace corresponding to the lowest input power is fitted with the nonlinear theory taking the Kerr into account to extract the actual photon number $n_{in}/n_{bi}=(0.54\pm0.02)$ and the Kerr constant $\mathcal{K}/2\pi = (14\pm1)\,\mathrm{kHz}$ of the setup.\\

With those parameters the associated photon number is extrapolated for each input power and the model is plotted to the experimental data. For each individual power, the cooling trace is well described by the extrapolated curves even though we increase the input power up to a factor of $6$ well beyond the bifurcation point. As soon as we reach $n_{in}/n_{bi} > 1$ two distinct branches for the cooling are observed in good agreement with the predictions of theory. However, it is evident that, except for the two sets where we measured with the pulse tube turned off, we cannot reach deep into the cooling features. This indicates, that despite the vibration isolation platform we still suffer from flux noise induced by mechanical vibrations in the system. Nevertheless, we have to emphasize, that for being that far beyond the bifurcation point the quality of the experimental data allows us to clearly demonstrate that backaction cooling and state manipulation in this regime are possible and predictable.  

%\begin{itemize}
%    \item fig with 0.5nbi to 5nbi: $n_{ph},\Gamma_m,\delta \omega_m$
%    \item good agreement
%    \item Goro cali still works?!
%    \item should we add right side fmap? 
%    \item should we show different g0?
%    \item explain gap?
%    \item explain drop?
%\end{itemize}

\noindent

%\noindent\begin{figure}[h]
%    \centering
%    \vspace{-100pt}
%    \includegraphics[height=0.8\paperheight]{chapters/figures_appendix/beyond_1dB_linfit_all_v2.pdf}
%    \caption{lin fit}
%    \label{fig:all_lin}
%\end{figure} 
\bibliography{cited_references}
%\end{CJK*}% Use default fonts from CJK (see below)
\end{document}